\newif\ifjournal
\journalfalse

\ifjournal
  \documentclass{aa}
  \usepackage{graphicx,times}
\else
  \documentclass{paper}
  \newcommand{\la}{\stackrel{{\scriptstyle <}}{{\scriptstyle \sim}}}
  \usepackage{graphicx,times}
\fi

\begin{document}

%\ifjournal \thesaurus{02.13.1, 11.03.1} \fi

\title{The temperature-mass relation in magnetized galaxy clusters}
\ifjournal
  \author{Klaus Dolag\inst{1} \and August Evrard\inst{2} \and
    Matthias Bartelmann\inst{1}}
  \offprints{K.~Dolag}
  \institute{Max-Planck-Institut f\"ur Astrophysik, P.O.~Box 1317,
    D--85741 Garching, Germany \and
    Dept. of Physics, University of Michigan, Ann Arbor, MI
    48109-1120, USA}
  \date{}
\else
  \author{Klaus Dolag$^1$, August Evrard$^2$, and Matthias
    Bartelmann$^1$\\
    $^1$Max-Planck-Institut f\"ur Astrophysik, P.O.~Box 1317, D--85741
    Garching, Germany\\
    $^2$Dept. of Physics, University of Michigan, Ann Arbor, MI
    48109-1120, USA}
  \date{}
\fi

\ifjournal
   \abstract{
\else
   \begin{abstract}
\fi
We use cosmological, magneto-hydrodynamic simulations of galaxy
clusters to quantify the dynamical importance of magnetic fields in
these clusters. The set-up of initial magnetic field strengths at high
redshifts is chosen such that observed Faraday-rotation measurements
in low-redshift clusters are well reproduced in the simulations. We
compute the radial profiles of the intracluster gas temperature and of
the thermal and magnetic pressure in a set of clusters simulated in
the framework of an Einstein-de Sitter and a low-density,
spatially-flat CDM cosmological model. We find that, for a realistic
range of initial magnetic field strengths, the temperature of the
intracluster gas changes by less than $\approx5\%$.
\ifjournal\fi
\ifjournal
   }
\else
   \end{abstract}
\fi

\maketitle

\section{Introduction}

Magnetic fields are common in galaxy clusters. They are inferred from
Faraday-rotation measurements (Vallee et al.~1986, 1987) in polarised
background radio sources, and from diffuse radio haloes around galaxy
clusters (Kronberg 1994). Depending on several assumptions, most
notably on the characteristic length scale of field coherence,
magnetic field strengths inferred from observations in typical galaxy
cluster centres range from $\approx0.1\mu{\rm G}$ up to some tens of
$\mu{\rm G}$.

Provided the magnetic field is tangled on sufficiently small scales
for it to be considered randomised, it contributes a non-thermal
pressure $p_{\rm B}$, which is related to the thermal pressure $p_{\rm
th}$ through
\begin{equation}
  \frac{p_{\rm B}}{p_{\rm th}} =
  % \frac{|{\mathbf B}|^2}{8\pi\,n\,kT} =
  2.5\%\,\left(\frac{|\vec{B}|}{3\mu{\rm G}}\right)^2\,
  \left(\frac{n}{10^{-3}\,{\rm cm}^{-3}}\right)^{-1}\,
  \left(\frac{T}{10^8\,{\rm K}}\right)^{-1}\;.
\label{eq:1}
\end{equation}
Conversely, magnetic fields ordered on scales comparable to the
cluster scale can add anisotropic non-thermal pressure.

The structure of intracluster magnetic fields is largely unclear at
present, and consequently the magnetic field strengths in clusters are
equally uncertain. We investigate in this paper the amount of
non-thermal pressure support due to realistic magnetic fields in
simulated galaxy clusters.

Our approach is as follows (see Dolag et al.~1999). We set up
gas-dynamical cosmological cluster simulations at high redshift and
impose magnetic fields of variable strength and structure as
summarised in Table~\ref{tab:bfield} below. Since the origin of
magnetic fields on cluster scales is unknown, we use two extreme
types of seed fields, namely either completely homogeneous or chaotic
initial magnetic field structures. The evolution of the intracluster
fields is then computed during cluster collapse. The initial 
mean magnetic field strength (or energy density) is a free
parameter in our simulations. We choose its range such that the
simulated clusters at low redshift well reproduce the statistics of
observed Faraday-rotation measurements taken from Kim et
al.~(1991). It turned out in earlier work (Dolag et al.~1999;
hereafter DBL99) that (i) the initial field structure is irrelevant
because the final field structure is completely determined by the
dynamics of the cluster collapse, and (ii) initial field strengths of
order $10^{-9}\,{\rm G}$ at the initial redshift of the simulations
(see Table \ref{tab:models}) are adequate to reproduce the measured
Faraday-rotation statistics.

\section{Models}

We use a set of 90 numerically simulated galaxy clusters for
investigating the effect of magnetic fields on the temperature and
pressure stratification in these clusters. Here, we summarise the
numerical technique only briefly because details on the methods
applied, their implementation, and their performance in test problems
have been published elsewhere (DBL99).

\subsection{Numerical method: GrapeMSPH}

We use the cosmological MHD code described in DBL99 to simulate the
formation of magnetised galaxy clusters from an initial density
perturbation field. The code combines the merely gravitational
interaction of a dark-matter component with the hydrodynamics of a
gaseous component. The gravitational interaction of the particles is
evaluated on GRAPE boards (cf.~Sugimoto et al.~1990), while the gas
dynamics is computed in the SPH approximation (Lucy 1977; Monaghan
1992). The original ``GrapeSPH'' code (Steinmetz 1996) was extended by
adding the magneto-hydrodynamic equations to trace the evolution of
the magnetic fields. Because of the assumed ideal electric
conductivity, the fields are frozen into the gas flow. The
back-reaction of the magnetic field on the gas via the Lorentz force
is included. The numerical viscosity required to capture shocks
in SPH is chosen such that angular-momentum transport in presence of
shear flows is carefully controlled. The SPH kernel width is
automatically adapted to the local number density of SPH particles,
which results in an adaptive spatial resolution of the code.

Extensive tests of the code were performed and described in DBL99.
The code succeeds in solving the co-planar MHD Riemann problem posed
by Brio \& Wu (1988). Although the simulated magnetic field is not
strictly divergence-free, $\nabla\cdot\vec{B}$ is always negligible
compared to the magnetic field divided by a typical length scale of
$\vec{B}$ (see DBL99 for details). The code also assumes the
intracluster medium to be an ideal gas with an adiabatic index of
$\gamma=5/3$, and neglects cooling. The surroundings of the clusters
are important because their tidal fields affect the overall cluster
structure and the merger history of the clusters. Therefore, the
cluster simulation volumes are surrounded by a layer of boundary
particles whose purpose it is to accurately represent the tidal fields
of the cluster neighbourhood. Here we specially focus on the dynamical
influence of the magnetic field in galaxy clusters, e.g.~on the
pressure support and the temperature in the clusters.

\subsection{Initial conditions}

We set up cosmological initial conditions for two CDM-dominated
universes (EdS and FlatLow) as specified in
Table~\ref{tab:models}. For each cosmology, we calculate ten different
realisations of the initial density-fluctuation field at redshift
$z_{\rm ini}$, which result in clusters of different final masses and
dynamical states at redshift $z=0$. We simulate each of these clusters
with up to five different initial magnetic field configurations,
listed in Table~\ref{tab:bfield}, yielding a total of 90 cluster
models.

\begin{table}[ht]
\caption{Parameters of the EdS and the FlatLow
models.\label{tab:models}}
\ifjournal\else\vspace{\baselineskip}\fi
\begin{center}
\begin{tabular}{|l|cccccc|}
\hline
 model & $H_0$ & $\Omega_0$ & $\Lambda$ & $\sigma_8$ &
 $f_\mathrm{baryon}$ & $z_\mathrm{ini}$ \\
\hline
 EdS     & 0.5 & 1.0 & 0.0 & 1.2  &  5\% & 15 \\
 FlatLow & 0.7 & 0.3 & 0.7 & 1.05 & 10\% & 20 \\
\hline
\end{tabular}
\end{center}
\end{table}

In lack of any detailed knowledge on the origin of primordial
seed fields, we explore two extreme cases of initial field
configurations. In one case (``homogeneous''), we assume that the
field is initially constant throughout the cluster volume. In the
other case (``chaotic''), we let the initial field orientation vary
randomly from place to place, subject only to the condition that
$\nabla\cdot\vec B=0$. The initial field strengths in both cases are
determined by setting the mean field energy
densities. Table~\ref{tab:bfield} summarises the initial field set-ups
and the mean field strengths in the cluster cores. The initial field
strengths are of order $10^{-9}\,$G, the final field strengths of
order $\mu$G.

\begin{table}[ht]
\caption{Initial magnetic fields (column 2) and the resulting final
magnetic field strengths in the simulated clusters (column 3 and 4 for
EdS and FlatLow models, respectively). The final values are an average
of the magnetic field over the central region (within a radius of
350$\,$kpc) and over all ten clusters for each
cosmology.\label{tab:bfield}}
\ifjournal\else\vspace{\baselineskip}\fi
\begin{center}
\begin{tabular}{|l|c|c|c|}
\hline
 initial field % model
 & $B_\mathrm{ini}$ 
 & $\left<B_\mathrm{final}\right>_\mathrm{core}^\mathrm{EdS}$ 
 & $\left<B_\mathrm{final}\right>_\mathrm{core}^\mathrm{FlatLow}$ \\
\hline
no
 & $0.0\,{\rm G}$              & ---               & ---\\
low (chaotic)
 & $0.2\times10^{-9}\,{\rm G}$ & $0.4\;\mu{\rm G}$ & ---\\
low
 & $0.2\times10^{-9}\,{\rm G}$ & $0.4\;\mu{\rm G}$ & $0.3\;\mu{\rm G}$\\
medium
 & $1.0\times10^{-9}\,{\rm G}$ & $1.1\;\mu{\rm G}$ & $0.8\;\mu{\rm G}$\\
high
 & $5.0\times10^{-9}\,{\rm G}$ & $2.5\;\mu{\rm G}$ & $2.0\;\mu{\rm G}$\\
\hline
\end{tabular}
\end{center}
\end{table}

\subsection{Evolution of intra-cluster fields}

At low redshifts, our simulated clusters reach temperatures of
$kT\la10\,{\rm keV}$, in agreement with observed clusters of similar
mass. The gas densities in their cores range within
$10^{-4}-10^{-3}\,{\rm cm}^{-3}$, as inferred from X-ray luminosities
and temperatures of observed clusters.

In agreement with our earlier paper (DBL99), we find that the
structure of the intra-cluster magnetic fields at low redshifts is
entirely independent on the structure of the initial magnetic
field. For homogeneous and chaotic initial field set-ups, the mean
strength and the coherence length are the same. Initial fields of
$10^{-9}\,{\rm G}$ at $z_{\rm ini}$ turn into final fields of $\mu{\rm
G}$ strength at redshift zero, and the typical coherence length of the
fields is of order $50\,{\rm kpc}/h$. In comparison to expectations
based on magnetic flux conservation in spherically collapsing spheres,
the amplification of the field is one order of magnitude higher
because of shear flows in the intracluster plasma.

\begin{figure*}[ht]
  \centerline{
    \includegraphics[width=0.49\hsize]{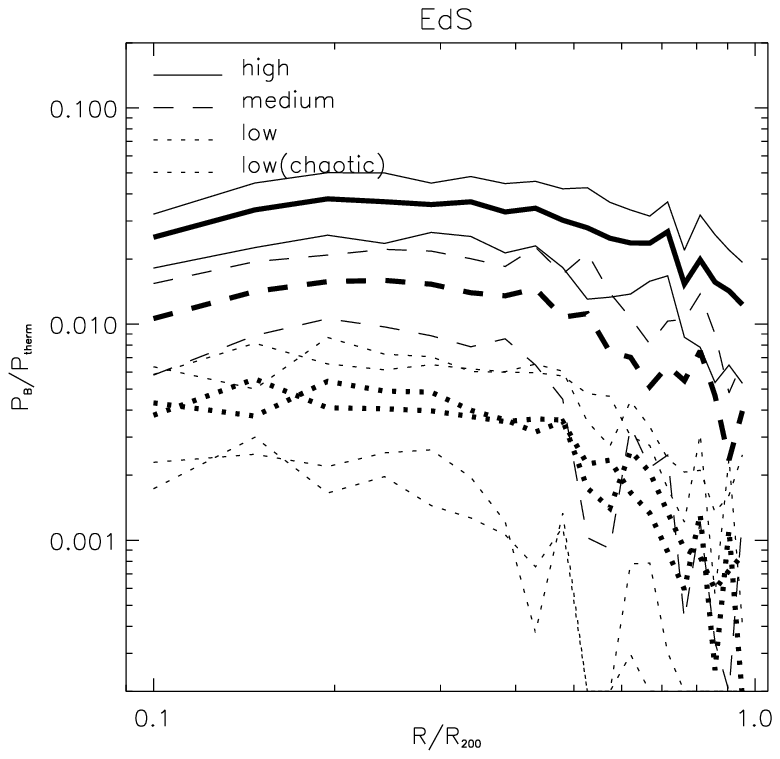}
    \includegraphics[width=0.49\hsize]{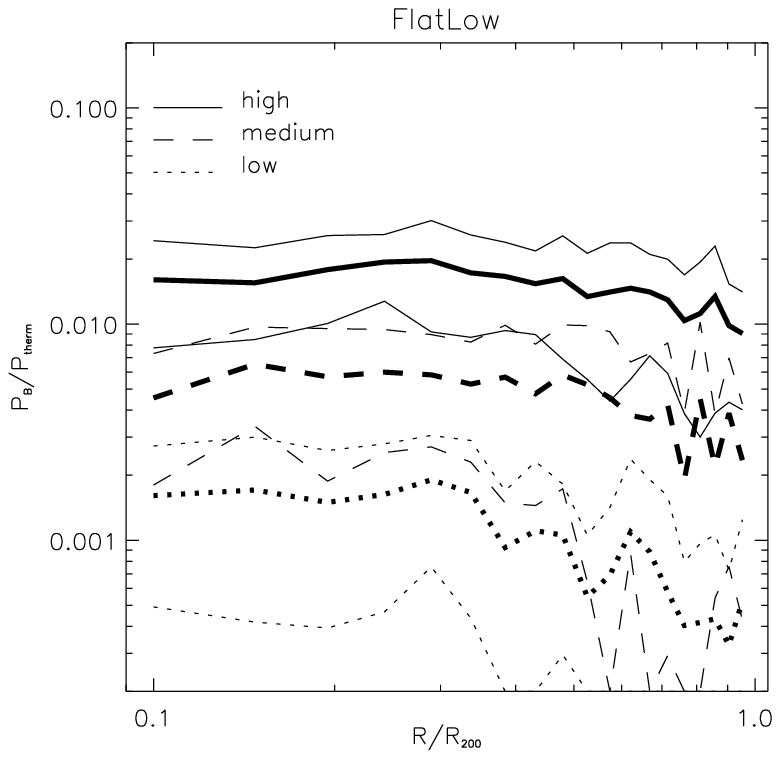}
  }
\caption{Shown here is the ratio between the magnetic and the thermal
pressure in shells around the cluster centre as a function of radius
in units of the virial radius $r_{200}$ (heavy lines). Different line
types distinguish the different initial magnetic field strengths as
indicated in the plot. The two dotted lines represent the low field
with homogeneous and with chaotic initial field configurations. Both
lead to the same amount of pressure support in the final clusters.
The curves show the average across the entire simulated cluster
sample. The thin lines indicate the {\em rms\/} scatter within the
sample. Near cluster centres, the ratio between magnetic and thermal
pressure is consistent with the isotropic estimate of
eq.~(\ref{eq:1}). The left panel represents the realisations for the
EdS cosmology, the right panel for the FlatLow cosmology.}
\label{fig:3}
\end{figure*}

During cluster collapse, the magnetic field is highly compressed,
twisted, tangled and sheared. Any information on its initial structure
is lost in that process. This is confirmed by Fig.~\ref{fig:3}, where
it is shown that the magnetic pressure support does not depend on the
initial field structure. We can therefore restrict our study to the
homogeneous initial field set-up, which is numerically easier to
handle.

Due to the violent motion of the intracluster plasma during cluster
collapse, the final magnetic field strength depends on the dynamics of
the cluster formation. Starting with identical initial magnetic field
set-ups, final field strengths differ by a factor of $4-5$, depending
essentially on cluster mass. The mean magnetic field strength drops by
about two orders of magnitude between the cluster centre and the
virial radius. A detailed study of the evolution and the structure of
intracluster magnetic fields in different cosmologies will be
described in a forthcoming paper.

A direct way to compare simulated magnetic fields in galaxy
clusters with observations is provided by Faraday-rotation
measurements. We created samples of synthetic rotation measures by
integrating the Faraday rotation along $10^4$ light rays randomly shot
through each of a sub-set of our cluster simulations. As in DBL99, the
cumulative rotation-measure distributions obtained in simulations with
either homogeneous or chaotic initial field set-up were statistically
identical, confirming once again that the structure of the initial
field has no observable consequences for the structure of the final
field processed by cluster collapse.

We further performed two types of comparison between synthetic and
observed Faraday-rotation data. In the first, we compared the radial
distribution of rotation measures seen in the Coma cluster (Kim et
al.~1990) with that obtained from one of our simulated clusters with
mass similar to Coma. In the second, the sample of rotation measures
compiled by Kim et al.~(1991) from observations of many Abell clusters
was compared to synthetic rotation measures obtained from our set of
simulated clusters. In both cases, synthetic rotation-measure
distributions were obtained along $10^4$ light rays randomly shot
through the simulations.

Both comparisons showed that the {\em medium\/} initial magnetic field
strength could reproduce the observed statistics very well. In both
cases, the simulated and the observed rotation-measure distributions
could not significantly be distinguished in a Kolmogorov-Smirnov
test. However, the scatter in both measurements and simulations is
large. This is due to the relatively low number of simulated clusters
per set, and to the selection criteria of galaxy clusters involved in
the simulations and the observations. We therefore decided to allow
for a range of initial magnetic field strengths which results in a
range of synthetic Faraday-rotation distributions which are still well
in agreement with the observed data. Further detail can be found
in DBL99 and Dolag (2000). Comparing with more recent data by Clarke
et al.~(1999), we find identical or even improved agreement. A study
of clusters in different cosmologies, using newer observational data,
is under way.

Choosing a realistic model to distribute relativistic electrons in the
simulated clusters, the simulations are also able to match the
properties of observed radio-haloes very well, including the observed
very steep relation between cluster temperature and radio luminosity
(see Dolag \& Ensslin 2000).

\begin{figure*}[ht]
  \centerline{
    \includegraphics[width=0.49\hsize]{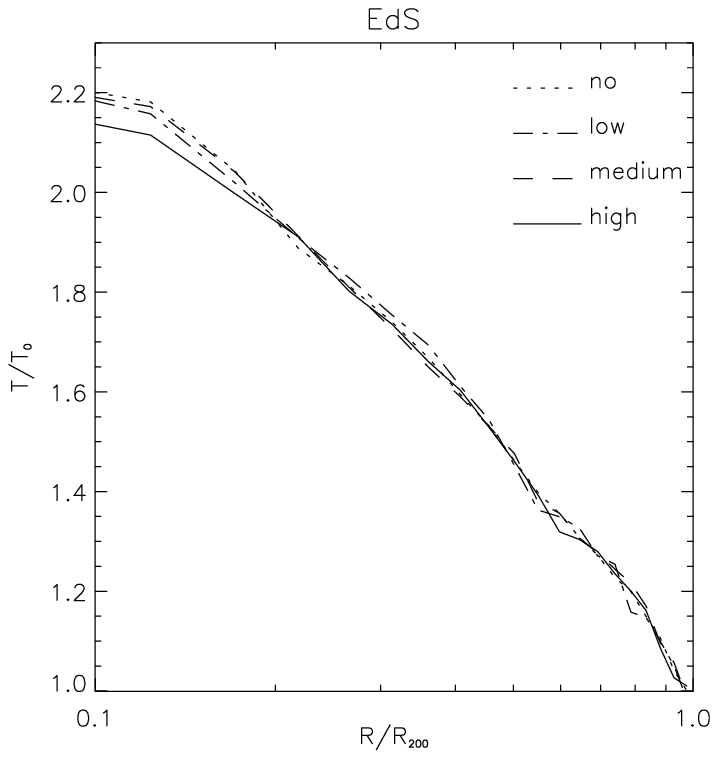}
    \includegraphics[width=0.49\hsize]{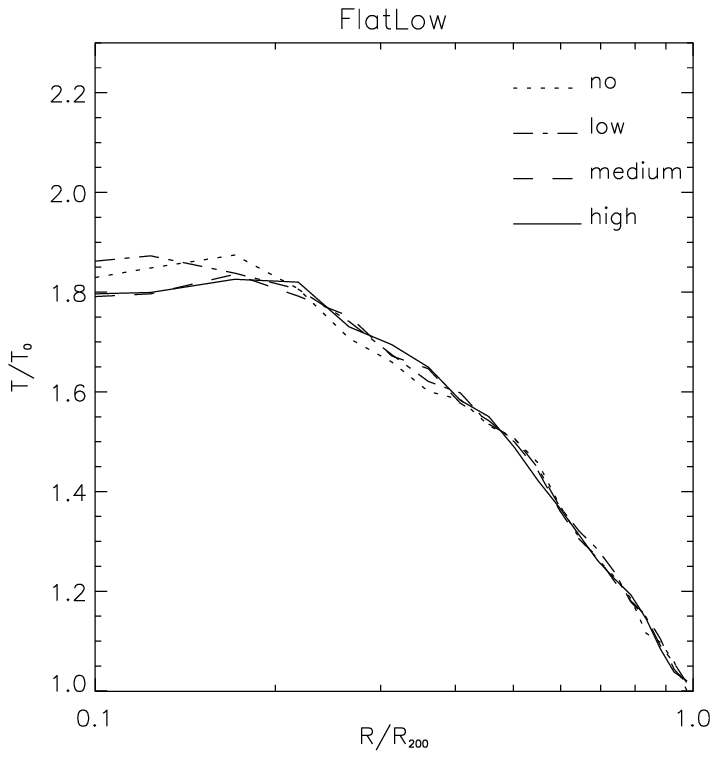}
  }
\caption{This figure shows the radial temperature profile, averaged
across the simulated cluster sample. The temperature is scaled by the
mass-weighted temperature of the non-magnetised cluster at a radius of
$r_{200}$, and the radius is scaled by $r_{200}$. Different line types
distinguish different initial magnetic field strengths, as indicated
in the plot. The temperature change with increasing magnetic field
strength is clearly visible. The temperature decrease near cluster
centres is compatible with the isotropic estimate of
eq.~(\ref{eq:1}). As in Fig.~\ref{fig:3}, the left panel represents
the EdS cosmology, the right panel the FlatLow cosmology.}
\label{fig:2}
\end{figure*}

\subsection{Quantities calculated}

The correlation length of final magnetic field configurations near
cluster centres is of order $50\,{\rm kpc}$ (see DBL99). Field
orientations in regions large compared to the correlation length can
therefore be considered as randomised. For instance, there is no
preferred direction in the final magnetic pressure distribution. Thus,
we can safely average thermodynamic quantities like the gas pressure
and gas temperature in spherical shells, provided these shells are
sufficiently thick compared to the field correlation length.

The radial distances to the cluster centres are scaled by the virial
radius $r_{200}$.  For simulated clusters, $r_{200}$ is defined as the
radius enclosing a mean density $\bar{\rho}(r_{200})$ of 200 times the
critical density $\rho_{\rm cr}$,
\begin{equation}
  \bar{\rho}(r_{200})=\frac{3M(r_{200})}{4\pi\,r_{200}^3}=
  200\,\rho_{\rm cr}\;.
\end{equation}
The temperature profiles give mass-weighted temperatures averaged in
shells, divided by the mass-weighted temperatures in the shell at the
radius $r_{200}$ of the corresponding non-magnetised cluster.

\section{Results}

During cluster formation and evolution, the simulated intracluster
magnetic field strengths are amplified by compression, tangling and
bending of field lines, and shear flows in the gas. The magnetic field
structure is also affected by merging, accretion of sub-clumps, and
consequent relaxation processes. In combination, these processes
amplify the initial magnetic fields in the simulation by factors of
order $\approx1000$. Final fields reach strengths of a few $\mu{\rm
G}$ in and near cluster centres, in good agreement with field
strengths inferred from observations. The growing magnetic fields
affect the balance between the gravitational force and the total
(magnetic plus thermal) pressure during cluster formation.

\begin{figure*}[ht]
  \centerline{
    \includegraphics[width=0.49\hsize]{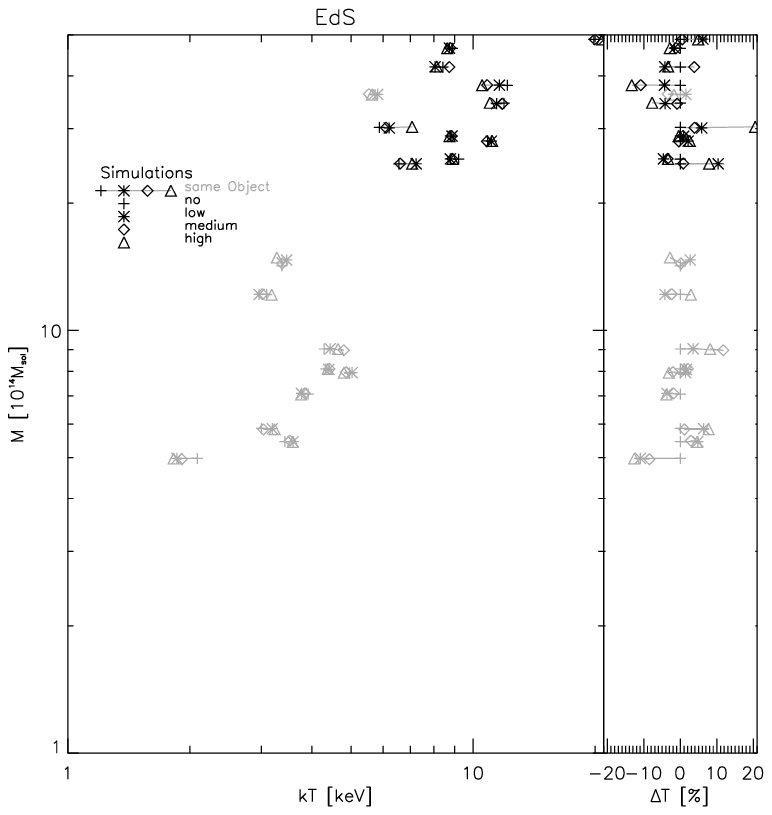}
    \includegraphics[width=0.49\hsize]{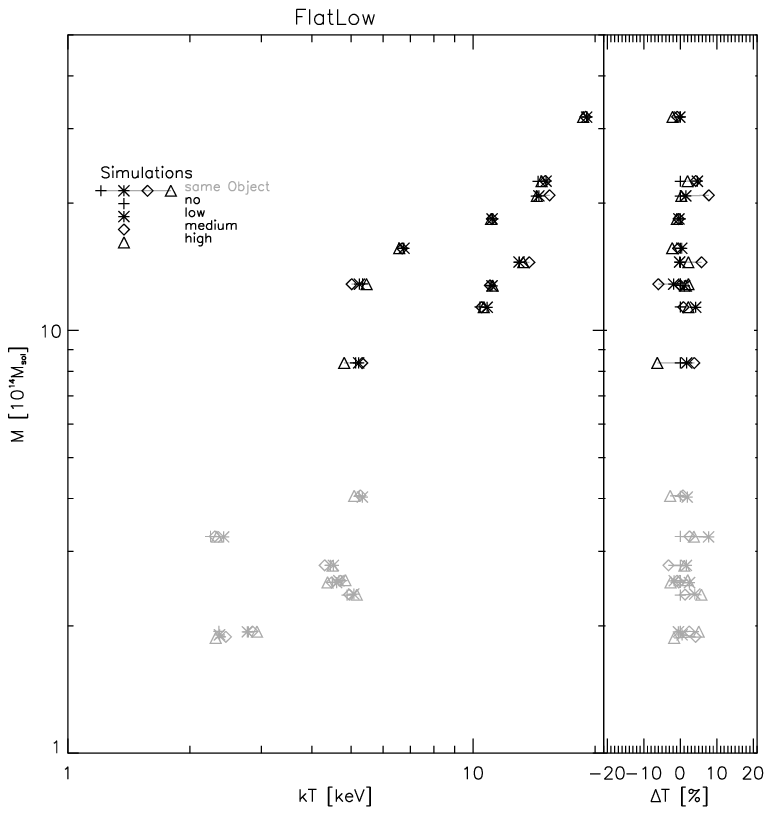}
  }
\caption{The large panels show the temperature-mass relation for the
simulated cluster sample. Here, the temperature is the
emission-weighted temperature within the radius $r_{200}$. All
clusters are simulated with different initial magnetic field strengths
marked by different symbols, as indicated in the plot. The right-hand
overlays show the temperature {\em change\/} relative to
non-magnetised clusters for the different initial magnetic field
strengths in percent for each simulated cluster. The large scatter is
obvious, but there is a general tendency towards lower temperatures in
magnetised clusters. To follow this relation towards lower
temperatures, smaller objects (marked as gray symbols) found in the
simulations are also used, which lead to a total of 20 different
clusters for the EdS cosmology (left panel) and 18 different clusters
for the FlatLow cosmology (right panel). As low mass objects are
represented by less particles, the resolution in these objects drops
towards lower masses and the scatter in these objects is expected to
be larger.}
\label{fig:1}
\end{figure*}

\subsection{Radial structure}

On the whole, the fairly isotropic magnetic pressure has the tendency
to counteract the gas collapse, and therefore acts in a similar manner
as the thermal pressure. This implies that the gas stratification can
achieve hydrostatic equilibrium at lower temperature, because lower
thermal pressure is required in presence of the additional pressure
support due to the magnetic field. Figure~\ref{fig:3} shows, averaged
in shells, the ratio
\begin{equation}
  {\cal F}_{\rm p}(R) =
  \frac{p_{\rm B}(R)}{p_{\rm th}(R)} =
  \frac
  {\left\langle|\vec{B}|^2/8\pi\right\rangle_{R_1\le R\le R_2}}
  {\left\langle p_{\rm th}\right\rangle_{R_1\le R\le R_2}}
\end{equation}
between the magnetic pressure $p_{\rm B}$ and the thermal pressure
$p_{\rm th}$ as a function of the radius $R$. Due to compression and
turbulence in the gas flow, the magnetic field is highest in and near
the centres of the simulated clusters, where the thermal pressure is
also highest. As Fig.~\ref{fig:3} shows, the relative contribution of
magnetic fields to the pressure support appears to peak somewhere
between the cluster core radii and the virial radii $r_{200}$. This
feature, however, is not very pronounced, so that the ratio between
magnetic and thermal pressure support is almost constant across a
large range of cluster-centric radii. Due to the smaller final masses
of the clusters in the FlatLow cosmology, the magnetic pressure
support is also smaller in these clusters. Nevertheless, these less
magnetized clusters are able to match the observed
Faraday-rotation. This is because the value of $H_0$ is chosen higher
in the low-density universe, so that the projected, physical distances
of the measurement locations with respect to the cluster centre
shrink. Closer to the cluster centre, the magnetic field is generally
higher, and therefore these less magnetised clusters well reproduce
the measurements.

The overall structure of the magnetic field in the outer regions of
the clusters differs from that near the cluster centre. Accretion and
merger events arrange the magnetic fields in the outer cluster regions
in coherent patterns on fairly large scales. Near the cluster centre,
the gas flow pattern is almost randomised, and the magnetic field is
consequently tangled and bent on fairly small scales. Although the
magnetic pressure can be considered isotropic at or near the cluster
centre, it may well be anisotropic further out. Even though the
fraction of the magnetic relative to the thermal pressure, averaged
within spherical shells, is approximately constant across the cluster,
the detailed effects of magnetic pressure on the gas flow depend on
the direction of the magnetic field relative to other locally
preferred directions, like the orientation of filaments surrounding a
cluster, the local path of matter infall, the orbit of a merging
sub-clump and such. Therefore, the same overall magnetic pressure can
have different effects on the local dynamics of gas flows, depending
on whether the magnetic field is ordered on scales comparable to the
cluster scale, and on the orientation of the magnetic field. In
addition, the magnetic field alters the dynamical time scale of a
(simulated) cluster. Therefore, when comparing simulated clusters with
and without magnetic fields at equal redshifts, we must be aware that
we may be comparing them at slightly different evolutionary stages
according to their intrinsic timescale.

Figure~\ref{fig:2} shows the temperature profile averaged across the
set of ten simulated clusters. On the whole, the temperature increases
towards the cluster centre by a factor of $\approx2$. For different
cosmologies, the final clusters have a different amount of
substructure. This leads to different shapes of the temperature
profiles for the clusters in the two different cosmologies used.
Although the detailed temperature profile of an individual cluster
depends on its dynamical state and formation history, the tendency is
clearly visible that the cluster temperature decreases with increasing
initial magnetic field strength towards the cluster centre. Deviations
from this general trend are possible in individual clusters because of
the reasons mentioned before. The mean across the entire cluster
sample, however, reveals a monotonic trend increasing towards the
cluster centre. This behaviour is expected in view of the previous
discussion about the impact of the magnetic field in different parts
of the cluster.

\subsection{Temperature-mass relation}

\begin{table}[ht]
\caption{This table shows the temperature difference between the
magnetised and non-magnetised clusters (columns 2 and 4), and the {\em
rms\/} (columns 3 and 5) of this distribution, for the three different
initial magnetic field strengths. Columns 2 and 3 are for the EdS,
columns 4 and 5 for the FlatLow cosmologies, respectively. The
temperatures are emission-weighted temperatures within the virial
radius $r_{200}$. All clusters shown in Figure~\ref{fig:3} are
used.\label{tab:dt}}
\ifjournal\else\vspace{\baselineskip}\fi
\begin{center}
\begin{tabular}{|l||c|c||c|c|}
\hline
 model & $\left<|\Delta T|\right>_\mathrm{EdS}$ 
       & {\em rms\/}
       & $\left<|\Delta T|\right>_\mathrm{FlatLow}$
       & {\em rms\/} \\
\hline
low   & 4.3\% & 2.7\% & 2.0\% & 2.1\%\\
medium& 3.1\% & 3.7\% & 2.4\% & 2.1\%\\
high  & 5.8\% & 5.1\% & 2.5\% & 1.7\%\\
\hline
\end{tabular}
\end{center}
\end{table}

The effect of magnetic fields on the temperature-mass relation in
galaxy clusters is shown in Fig.~\ref{fig:1}. Different initial
magnetic field strengths are marked with different symbols. As shown
before, the effect of the magnetic pressure on the temperature is on
average strongest near cluster centres. Therefore, the strength of the
effect on the temperature-mass relation depends on the chosen radius
within which the mean temperature is calculated. It changes also if we
choose the emission-weighted instead of the mass-weighted
temperature. For Fig.~\ref{fig:1}, we have chosen to calculate the
emission-weighted temperature within the virial radius $r_{200}$ in
order to be comparable to the measurements. We also translated the
masses of the simulated galaxy clusters to velocity
dispersions. Figure~\ref{fig:1} also shows the measured relationship
between cluster temperatures and velocity dispersion as dashed-dotted
line. The fairly large scatter in individual clusters is clearly
seen. In the mean, the temperature difference between the magnetised
and non-magnetised clusters is $\approx5\%$, and it can go up to
$10-15\%$ in individual clusters. 

Although the systematic effect of magnetic fields on the
emission-weighted temperature profiles is small, the magnetic fields
also increase the scatter in the mass-temperature relation.  Values
for the {\em rms\/} scatter are given in Table~\ref{tab:dt}.  This
effect is more pronounced in the EdS cosmology because of the smaller
gas fraction in a high-density universe.

\section{Discussion}

We tested the influence of the non-thermal pressure support due to
magnetic fields in galaxy clusters. To do so, we performed
cosmological MHD simulations of ten clusters in two different
cosmologies, each with four different initial magnetic field
configurations, thus yielding a total of 90 cluster models.

We chose initial magnetic field strengths such that the final
intracluster fields were able to statistically reproduce the amplitude
and spatial distribution of observed Faraday-rotation
measurements. For such fields, we found a relatively small
non-thermal pressure support of about $5\%$. The temperature change in
the cluster cores is therefore on average $\approx5\%$, with the
general tendency towards decreasing temperature with increasing
magnetic field strength. This tendency is most pronounced in
cosmologies with a low gas fraction, i.e.~a high dark-matter
density. The effect of the magnetic fields can be substantially
stronger in individual clusters, introducing a cosmology-dependent
scatter of up to $(10-15)\,\%$ into the mass-temperature relation.

Our overall conclusion is that the statistical rotation-measure
data, taken in the most straightforward interpretation, suggests that
magnetic fields are generally dynamically unimportant in galaxy
clusters. We are aware that simulations like those presented here do
not include the full range of physical processes required for a
definitive calculation of intracluster fields, like small-scale
turbulence and non-ideal magneto-hydrodynamics. A detailed study of
such processes is currently out of reach for realistic numerical
simulations, rendering our simulations illustrative rather than
definitive. We point out, however, that the inclusion of magnetic
fields in cluster simulations does constitute an improvement over
previous studies, even if it confirms earlier conjectures.

\ifjournal
   \acknowledgements{
\else
   \section{acknowledgements}
\fi
We wish to thank Harald Lesch for useful
discussions, and the referee, Jean Eilek, for her very detailed and
thoughtful report.
\ifjournal
   }
\fi

\end{document}